%% file: main.tex
\documentclass[final,3p,times,authoryear]{elsarticle}

\usepackage[T1]{fontenc}
\usepackage{mathptmx}
\usepackage{graphicx}
\usepackage{amssymb}
\usepackage{booktabs}
\usepackage{array}
\usepackage{lineno}
\usepackage[hidelinks]{hyperref}

\makeatletter
\def\ps@pprintTitle{\let\@oddhead\@empty\let\@evenhead\@empty
  \def\@oddfoot{\hfill\thepage\hfill}\let\@evenfoot\@oddfoot}
\makeatother

\newif\ifanon
\anonfalse  

\journal{Global Environmental Change}

\begin{document}

\begin{frontmatter}

\title{A System-of-Systems Convergence Paradigm for Societal Challenges of the Anthropocene}

\ifanon\else
\author[vt]{Megan S. Harris\corref{cor1}}
\ead{megan_harris@ncsu.edu}
\author[vt]{Mohammad Mahdi Naderi}
\author[sit]{Ehsanoddin Ghorbanichemazkati}
\author[fsu]{Sina Jangjoo}
\author[ub]{Emily Lapan}
\author[sit]{Amirreza Hosseini}
\author[iiasa]{Fabian Schipfer}
\author[csiro]{Stephen Craig}
\author[csiro]{Enayat Moallemi}
\author[wsu]{Inas Khayal}
\author[ub]{Laura M. Arpan}
\author[fsu]{Tian Tang}
\author[vt]{John C. Little}
\author[sit]{Amro M. Farid}

\cortext[cor1]{Corresponding author.}

\affiliation[vt]{organization={Department of Civil and Environmental Engineering, Virginia Tech},
  city={Blacksburg}, state={VA}, postcode={24061}, country={United States}}
\affiliation[sit]{organization={School of Systems and Enterprises, Stevens Institute of Technology},
  city={Hoboken}, state={NJ}, postcode={07030}, country={United States}}
\affiliation[fsu]{organization={Askew School of Public Administration and Policy, Florida State University},
  city={Tallahassee}, state={FL}, postcode={32306}, country={United States}}
\affiliation[ub]{organization={Department of Environment and Sustainability, University at Buffalo},
  city={Buffalo}, state={NY}, postcode={14260}, country={United States}}
\affiliation[iiasa]{organization={Biodiversity and Natural Resources Program, International Institute for Applied Systems Analysis (IIASA)},
  addressline={Schlossplatz 1}, postcode={A-2361}, city={Laxenburg}, country={Austria}}
\affiliation[csiro]{organization={Commonwealth Scientific and Industrial Research Organisation (CSIRO)},
  addressline={Private Bag 10}, city={Clayton South}, state={VIC}, postcode={3169}, country={Australia}}
\affiliation[wsu]{organization={Departments of Oncology, Industrial and Systems Engineering, and Computer Science, Wayne State University},
  city={Detroit}, state={MI}, postcode={48201}, country={United States}}
\fi

\begin{abstract}
Climate, water, food, energy and governance challenges are interdependent, yet the tools used to study and manage them remain divided by discipline and sector. Coupling sectoral models pair by pair cannot keep pace with the cross-sector and institutional interactions that drive global environmental change. In this Perspective we argue that convergence requires a shared system ontology that preserves meaning as knowledge moves from observation to computation and back to decision-making. Building on a meta-cognition map of scientific knowledge generation, we propose six design criteria for convergent system-of-systems approaches and assess four candidate methods against them, identifying model-based systems engineering paired with hetero-functional graph theory as a candidate ``convergence spine.'' We illustrate the paradigm in the Chesapeake Bay Watershed, where a single ontology links nutrient-flow estimation, institutional analysis of restoration governance, team science and graduate education, and we distinguish what has been implemented from what remains future work. We close with implications for how global environmental change research is organized, funded and taught.
\end{abstract}

\begin{highlights}
\item Coupling sectoral models pair by pair cannot capture cross-sector interactions
\item Convergence needs a shared ontology from observation through computation
\item Six design criteria are proposed for convergent system-of-systems approaches
\item SysML with hetero-functional graph theory is assessed as a convergence spine
\item Chesapeake Bay work links watershed models, governance analysis and teaching
\end{highlights}


\end{frontmatter}

\input{0-fullpaper}

\ifanon\else
\section*{CRediT authorship contribution statement}
\textbf{Megan S. Harris:} Conceptualization, Writing -- original draft, Writing -- review \& editing, Visualization.
\textbf{Mohammad Mahdi Naderi:} Writing -- original draft, Writing -- review \& editing, Visualization.
\textbf{Ehsanoddin Ghorbanichemazkati:} Writing -- original draft, Writing -- review \& editing.
\textbf{Sina Jangjoo:} Writing -- original draft, Writing -- review \& editing.
\textbf{Emily Lapan:} Writing -- original draft, Writing -- review \& editing.
\textbf{Amirreza Hosseini:} Writing -- original draft, Writing -- review \& editing.
\textbf{Fabian Schipfer:} Conceptualization, Writing -- review \& editing.
\textbf{Stephen Craig:} Conceptualization, Writing -- review \& editing.
\textbf{Enayat Moallemi:} Conceptualization, Writing -- review \& editing.
\textbf{Inas Khayal:} Conceptualization, Writing -- review \& editing.
\textbf{Laura M. Arpan:} Supervision, Funding acquisition, Writing -- review \& editing.
\textbf{Tian Tang:} Supervision, Funding acquisition, Writing -- review \& editing.
\textbf{John C. Little:} Conceptualization, Supervision, Funding acquisition, Writing -- review \& editing.
\textbf{Amro M. Farid:} Conceptualization, Supervision, Funding acquisition, Writing -- review \& editing. 
\fi

\section*{Declaration of competing interest}
The authors declare that they have no known competing financial interests or personal relationships that could have appeared to influence the work reported in this paper.

\section*{Data availability}
No new data were created or analyzed in this Perspective.

\ifanon\else
\section*{Acknowledgments}
This research is based on work supported by the Growing Convergence Research Program of the National Science Foundation under Grant Numbers OIA 2317874 and OIA 2317877.
\fi

\section*{Declaration of generative AI and AI-assisted technologies in the manuscript preparation process}
During the preparation of this work the authors used Claude (Anthropic) to assist in condensing the manuscript text. After using this tool, the authors reviewed and edited the content as needed and take full responsibility for the content of the published article.

\bibliographystyle{elsarticle-harv}
\bibliography{LIINESLibrary,LIINESPublications,convergenceLIINESreferences}

\end{document}

%% file: 0-fullpaper.tex
\section{Introduction}\label{sec:intro}
\begin{figure}[h!]
    \centering
    \includegraphics[width=0.98\linewidth]{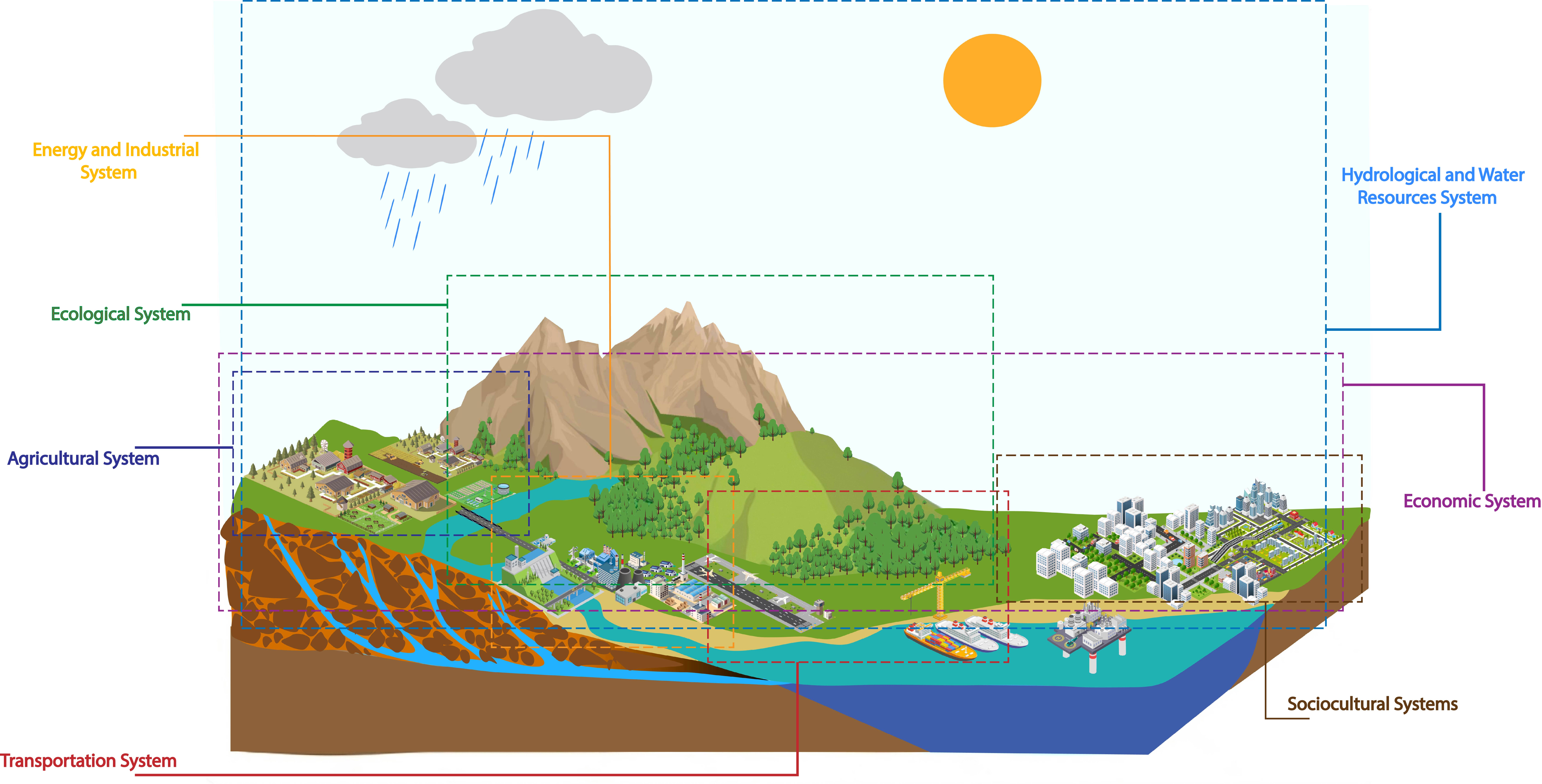}
    \caption{Graphical illustration of an interconnected geophysical, biophysical, sociocultural and sociotechnical system of Anthropocene systems.}
    \label{fig:CHNS}
\end{figure}

The Anthropocene era is marked by profound and irreversible human-driven transformations of Earth's natural and built systems~\citep{williams:2022:00, steffen:2011:00}.
Climate change, eutrophication, food insecurity, water scarcity and energy insecurity are not isolated crises; they are interwoven and mutually reinforcing \citep{steffen:2011:00,ho:2019:00,stavi:2021:00}. These challenges span coupled geophysical, biophysical, sociocultural and sociotechnical systems (Fig.~\ref{fig:CHNS}; \citealp{Little:2023:ISC-J50}), yet the institutions governing them rarely align with the scope of the problems they address~\citep{Cornell:2013:00,berkes:2017:00}. Progress in one sector can undermine another: regions coping with water scarcity often turn to energy-intensive supply options that raise greenhouse gas emissions, and decarbonization pathways can increase demands on land and water \citep{kraucunas:2015:00,Bi:2022:00}.

Integrated assessment and socio-environmental modeling have made real progress in linking these systems~\citep{Iwanaga:2021:00,razavi:2025:00,yang:2025:00}. However, responses in research, policy and practice remain largely organized around single-discipline, single-problem approaches, supported by siloed institutions and legacy analytical tools~\citep{verburg:2016:00}. Even ambitious nexus studies often fall short: a systematic review of 245 food--energy--water publications found that most did not capture all three sectors, let alone their interactions~\citep{Albrecht:2018:00}. The underlying hurdle is not a shortage of models, but the lack of a shared language for relating models, data and institutions.

 \citet{Little:2026:ISC-JR10} recently proposed an evolutionary system-of-systems (SoS) convergence paradigm and outlined its requirements: modular, hierarchical conceptual models grounded in a reconciled ontology, extensible computational frameworks, decision-support systems, and educational pedagogy. This Perspective asks the next question: \emph{how can multi-discipline, multi-problem convergence be built and evaluated in practice?} We argue that convergence requires more than coupling increasingly specialized sectoral models; it requires a shared system ontology that preserves meaning across observation, systems conceptualization, visual representation, mathematics, and computation. This Perspective makes three contributions. First, it employs a meta-cognition map to illustrate how meaning can be lost during knowledge generation without a shared ontology. Second, it proposes six design criteria for convergent SoS approaches and evaluates four candidate methods against them. Third, it demonstrates the resulting paradigm through a Chesapeake Bay Watershed case study linking computational modeling, governance analysis, team science, and education. The case study explicitly distinguishes what has been achieved from what remains future work.


\section{Common Pitfalls to Convergence}\label{sec:pitfalls}

Convergence is hindered by epistemic, institutional, and infrastructural barriers, not merely technical limitations \citep{razavi:2025:00}. Drawing on prior research and experiences from convergence science initiatives, we identify five systemic pitfalls.

\emph{Regional and institutional context dependence.} Tools developed for one region or institutional setting seldom transfer directly to another~\citep{yates:2018:00}. The Chesapeake Bay's nutrient accounting framework works under total maximum daily load (TMDL) mandates and strong interstate coordination~\citep{ferris:2024:00}, conditions that differ from those in the Mississippi Basin, where agricultural regulation and coordination are weaker~\citep{secchi:2019:00}. Convergent approaches must distinguish generalizable from context-specific system structure.

\emph{Fragmented ontologies and sectoral silos.} Tools are often developed within individual sectors, each with its own definitions, data structures, and modeling assumptions~\citep{Farid:2022:01,reed:2022:01}. These differences complicate integration and can produce unrealistic cross-sector assumptions, such as decarbonization scenarios that pair detailed power-sector models with coarse or static representations of water availability or bioenergy feedstock supply~\citep{kraucunas:2015:00}. The IPCC notes that maladaptation can be avoided through flexible, multi-sectoral, inclusive and long-term planning~\citep{ipcc:2023:01}, but such planning presupposes that sectors can describe their systems in mutually intelligible terms.

\emph{Bottom-up modeling misses cascading interdependencies.} Disciplines often model their systems independently before attempting to couple them. Although tractable, this approach can miss cascading effects across climate, agriculture, energy, land use, and governance. For example, drought in the western United States can reduce hydropower output, increase reliance on natural-gas generation, and raise emissions~\citep{bartos:2015:00}. The combinatorial challenge compounds the problem: ten sectors with 100 nested systems each yield more than one million potential cross-sector interactions. Even if only 1\% are relevant, identifying and characterizing them individually would be impractical~\citep{Bi:2022:00,Farid:2022:01}. These limitations motivate combining top-down system-wide design with bottom-up process-level modeling. 

\emph{Co-simulation limitations in structure and scale.} Co-simulation commonly connects models through sequential exchanges of inputs and outputs~\citep{gomes:2017:00}, which can complicate the representation of multilateral feedbacks and distributed decision-making across systems~\citep{taveres-Cachat:2021:01}. It also faces challenges when coupled processes operate on different timescales, such as decadal land-use and infrastructure decisions alongside hourly operations~\citep{turner:2017:00}. Convergent modeling must therefore accommodate both cross-system feedbacks and interactions across timescales.

\emph{Overlooking social and institutional dynamics.} Technical frameworks often omit trust, public values, cultural narratives, and governance because they are treated as outside the scope of technical analysis~\citep{van-bruggen:2019:00,loorbach:2017:00}. Opposition to wind farms, for example, can reflect concerns about fairness, place attachment, and procedural exclusion rather than technology performance alone~\citep{devine-wright:2011:00,wolsink:2018:00}. When these dynamics are excluded from modeling and stakeholder engagement, technically sound interventions may lack legitimacy or public acceptance~\citep{cash:2003:01}. Convergent approaches must therefore account for the social and institutional conditions that shape system change.

Together, these pitfalls reveal a common structural deficiency: the lack of an integrated scientific and institutional infrastructure for connecting disciplinary models, data, and governance. Addressing this gap requires a use-inspired convergence paradigm grounded in a shared ontology and extensible across scales, sectors, and regions, with methods and educational tools co-developed alongside stakeholders. 

\section{Meta-Cognition Map as the Scaffold for a Convergent SoS Paradigm}\label{sec:metacognition}

\begin{figure}[t]
  \centering
  \includegraphics[width=\textwidth]{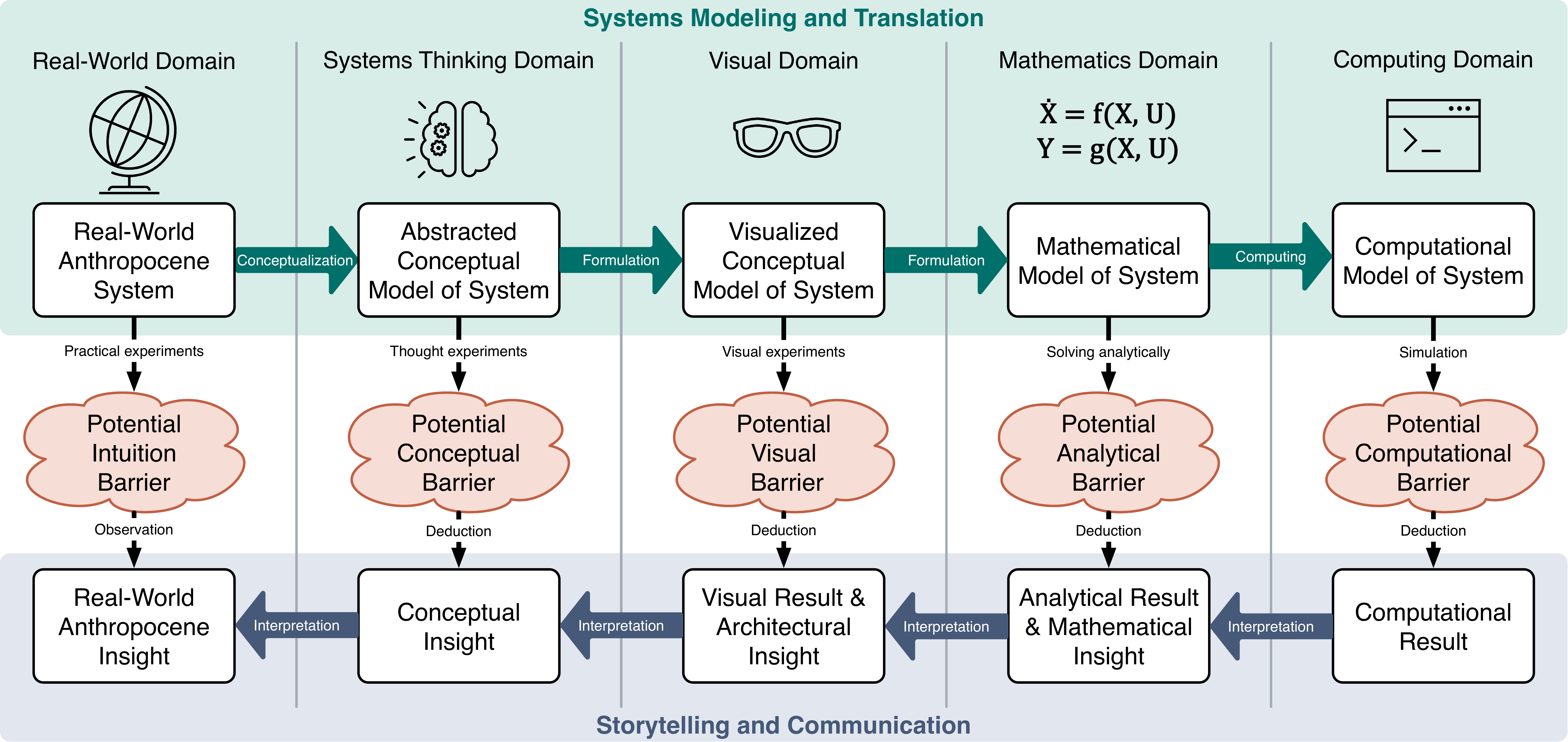}
  \caption{Meta-cognition map of scientific knowledge generation. Researchers move among five complementary domains (Real-World, Systems Thinking, Visual, Mathematics, Computing). Conceptualization, formulation and computing carry knowledge forward (top); experiments within each domain encounter domain-specific barriers (middle); interpretation carries results back toward real-world insight (bottom). A shared ontology is what keeps meaning intact across each translation. Adapted from \citet{Farid:2023:01} and \citet{Little:2026:ISC-JR10}.}
  \label{fig:metacog}
\end{figure}

The meta-cognition map (Fig.~\ref{fig:metacog}) describes how researchers generate cross-domain knowledge as they move among five domains~\citep{Farid:2023:01,Little:2026:ISC-JR10}. Inquiry begins in the \emph{Real-World} domain, through observation, sensing, and experiment; for global systems, controlled experiments are rarely possible, and observations are incomplete and noisy~\citep{reichstein:2019:00}. \emph{Systems thinking} synthesizes observations into mental models of interconnection, feedback, and emergence, but these subjective models lack formal structure \citep{forrester:1990:00}. The \emph{Visual} domain externalizes mental models as diagrams, maps, and architectures that collaborators can inspect \citep{mann:2009:00}. \emph{Mathematics} provides rigor and inference, but abstractions may simplify nonlinearity, uncertainty, and context~\citep{beltrami:2014:00}. \emph{Computing} operationalizes mathematical models for high-dimensional problems, but its results are highly sensitive to upstream assumptions \citep{saltelli:2020:00}.

Each domain produces insight and carries its own barrier.
Knowledge is produced by moving forward through the domains by conceptualization, formulation, and computing as insights are redefined and re-expressed, maintaining ontological coherence \citep{Guizzardi:2005:01}. Insights then traverse backward through interpretation, so that computational results become mathematical, architectural, and conceptual and, ultimately, guidance for real-world intervention. The map makes visible where convergence fails: at each translation. When an ecologist's ``nutrient load,'' an economist's ``output'' and a regulator's ``credit'' are formalized separately, meaning is lost or distorted with each hand-off. 
A shared ontology, or common set of fundamental system elements and relations used consistently across representations~\citep{Guizzardi:2005:01}, allows knowledge to cross these boundaries intact.

The return path matters as much as the forward one. Results must travel beyond the Computing domain, back through the mathematical, visual, and conceptual domains, to become narratives that decision-makers and communities can evaluate and act on (Fig.~\ref{fig:metacog}, bottom). A shared ontology supports this translation by allowing the same system elements that structure a model to structure its explanation, keeping the assumptions behind recommendations traceable to those affected by them~\citep{cash:2003:01}.

This framing has practical consequences. Integration efforts often focus on the Mathematics and Computing domains by building larger models and faster couplers, while leaving upstream translations informal. Recent work in healthcare delivery suggests that, in complex sociotechnical settings, the central challenge often lies less in the mathematics than in establishing shared mental models and system-level framing~\citep{Satcher:2025:ISC-JR05}. Convergence therefore requires methods that connect the domains, not merely better tools within any one of them.

\section{Design Criteria and Candidate Approaches}\label{sec:criteria}

\begin{figure}[t]
  \centering
  \includegraphics[width=\textwidth]{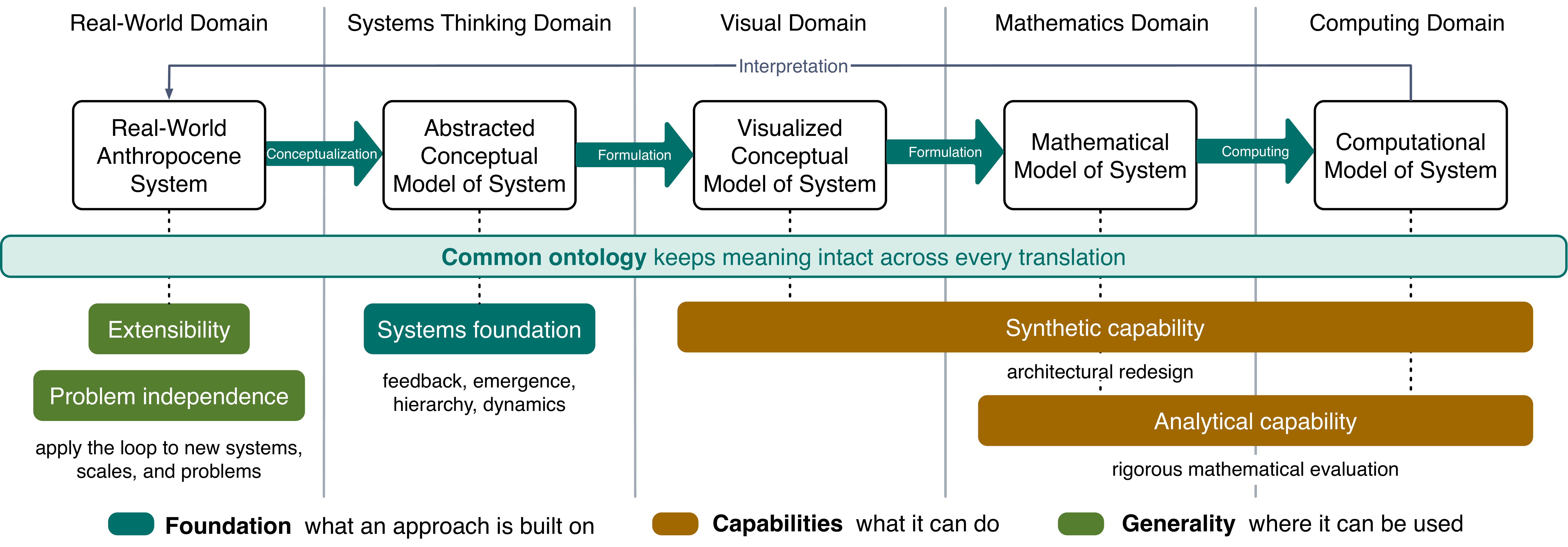}
  \caption{Six design criteria for convergent system-of-systems approaches, mapped onto the meta-cognition map. A common ontology spans every translation between domains; the remaining criteria are grouped by foundation (systems foundation), capabilities (analytical and synthetic) and generality (extensibility and problem independence).}
  \label{fig:criteria}
\end{figure}

Building on the meta-cognition map and on an earlier proposal to assess the convergence potential of systems-engineering methods \citep{Farid:2022:01}, we propose six design criteria (Fig.~\ref{fig:criteria}). Each responds to one or more of the barriers in Section~\ref{sec:pitfalls}.
\begin{enumerate}
  \item A \emph{systems foundation} that captures feedback, emergence, hierarchy and dynamics, allowing physical and social systems to be integrated \citep{forrester:1990:00,weck:2011:00}.
  \item A \emph{common ontology} that keeps visual, mathematical and computational representations consistent and translatable across disciplines \citep{Guizzardi:2005:01,dori:2016:01}.
  \item \emph{Extensibility}, so that new systems, sectors and scales can be added without structural redesign \citep{Little:2019:00}.
  \item \emph{Analytical capability} for rigorous mathematical evaluation of system behavior.
  \item \emph{Synthetic capability} for architectural design and redesign, not only analysis of what exists \citep{dori:2016:01}.
  \item \emph{Problem independence}, so that the same approach serves water, energy, food, health or governance questions in different regions \citep{yates:2018:00}.
\end{enumerate}

\begin{table}[t]
\centering
\footnotesize
\caption{Qualitative assessment of four candidate methods against the six design criteria ($\bullet$ strong; $\circ$ partial; -- weak or absent). Assessment by the authors, based on the discussion in Section~\ref{sec:criteria} and \citet{Farid:2022:01}.}
\label{tab:methods}
\begin{tabular}{@{}l>{\centering\arraybackslash}p{1.45cm}>{\centering\arraybackslash}p{1.45cm}>{\centering\arraybackslash}p{1.45cm}>{\centering\arraybackslash}p{1.45cm}>{\centering\arraybackslash}p{1.45cm}>{\centering\arraybackslash}p{1.45cm}@{}}
\toprule
Method & Systems foundation & Common ontology & Extensibility & Analytical & Synthetic & Problem independence \\
\midrule
Network science        & $\circ$   & --        & $\bullet$ & $\bullet$ & $\circ$   & $\bullet$ \\
Data-driven AI         & --        & --        & $\bullet$ & $\bullet$ & $\bullet$ & $\bullet$ \\
MBSE (SysML)           & $\bullet$ & $\bullet$ & $\bullet$ & $\circ$   & $\bullet$ & $\bullet$ \\
HFGT                   & $\bullet$ & $\bullet$ & $\bullet$ & $\bullet$ & $\bullet$ & $\bullet$ \\
\bottomrule
\end{tabular}
\end{table}

We assess four methods with high convergence potential against these criteria (Table~\ref{tab:methods}). \emph{Network science} offers a general visual and mathematical language for connectivity, resilience and modularity, and scales well, but its homogeneous nodes and edges struggle to represent the heterogeneity of real socio-environmental systems \citep{weck:2011:00,Farid:2022:01}. \emph{Data-driven artificial intelligence} excels at finding patterns in large, noisy data and can both diagnose and propose designs, but it carries no inherent systems ontology and risks black-box results that obscure causal understanding unless anchored to an ontologically grounded framework \citep{reichstein:2019:00,saltelli:2020:00}. \emph{Model-based systems engineering} (MBSE), using the Systems Modeling Language (SysML), represents system structure, behavior and requirements in a standardized, domain-neutral graphical language that spans the Systems Thinking, Visual and Mathematics domains and supports traceable translation between them \citep{mann:2009:00,dori:2016:01}; on its own, however, it has limited analytical power. \emph{Hetero-functional graph theory} (HFGT) models systems through their functional capabilities, the processes by which resources transform, store and transport operands, rather than through component identities alone, yielding a mathematically rigorous, ontology-driven representation that scales to large computational problems \citep{Farid:2022:01,Farid:2025:ISC-JR06}.

The four are complementary. Network science abstracts connectivity, AI extracts knowledge from data, MBSE structures coherent architectures, and HFGT turns those architectures into analyzable and computable form. We identify SysML-based MBSE paired with HFGT as a candidate ``convergence spine'': SysML supplies the graphical ontology and architectural traceability, HFGT the mathematical ontology and computational rigor. Over the past decade this pairing has been applied to power grids, water distribution, transport and manufacturing, and to systems of systems including hydrogen--natural gas infrastructure, multi-modal energy systems and distributed healthcare \citep{Khayal:2018:00,Schoonenberg:2022:ISC-J48,Thompson:2024:ISC-J53}. A formal comparison with system dynamics, long central to human--natural systems analysis, showed that SysML and HFGT encode a broader set of systems-thinking constructs, reproduce system-dynamics results and remove barriers to multi-domain integration \citep{Naderi:2026:ISC-JR08}. We present this spine as a candidate to be tested, not as a finished solution.

\section{The paradigm in the Chesapeake Bay Watershed}\label{sec:case-study}

\begin{figure}[t]
  \centering
  \includegraphics[width=\textwidth]{\ifanon figures/casestudyanon.pdf\else 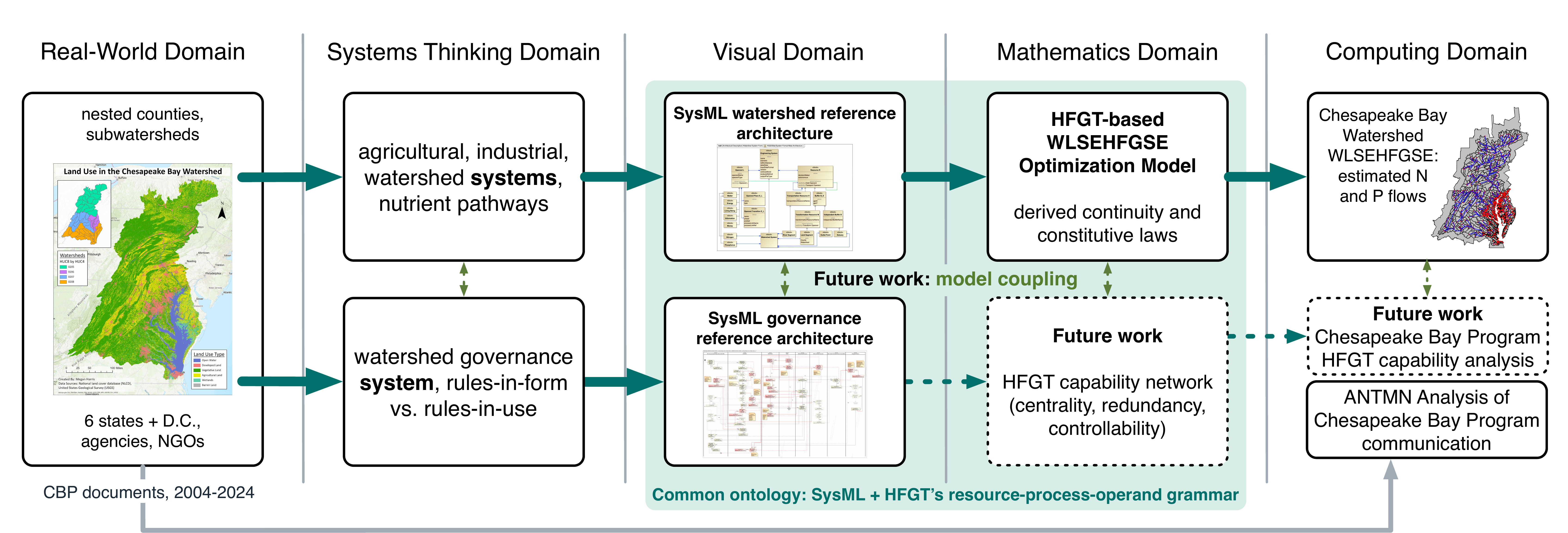\fi}
  \caption{The paradigm instantiated for the Chesapeake Bay Watershed and its governance. The watershed system (top) and the governance system (bottom) share a SysML--HFGT ontology. Solid boxes indicate implemented components; dashed boxes indicate future work. The ANTMN analysis of Chesapeake Bay Program committee documents (2004--2024) links real-world governance records to computational analysis. Architectures and river-network map from \citet{Harris:2026:01} and \citet{Jangjoo:2026:ISC-JR09}; land use from NLCD/USGS.}
  \label{fig:casestudy}
\end{figure}

The Chesapeake Bay Watershed concentrates the hallmark difficulties of global environmental change in a single region. It spans six states and the District of Columbia, drains more than 100,000 tributaries into the largest estuary in the United States, and supports over 17 million people \citep{pan:2021:00}. Restoration is coordinated by the Chesapeake Bay Program (CBP), a partnership of federal and state agencies, local governments, non-governmental organizations and universities, to address persistent hypoxia, eutrophication and habitat loss \citep{Hood:2021:00,batiuk:2025:00}. The CBP's modeling system, including the Chesapeake Assessment Scenario Tool (CAST), is scientifically sophisticated but difficult to extend and structurally opaque to many participants \citep{Hood:2021:00}. The region thus combines scientific complexity, institutional fragmentation and socio-economic coupling. 
\ifanon
A multi-institution convergence research project has used it as a testbed
\else
The \textbf{Convergent} \textbf{Anth}ropocene Syst\textbf{ems} (\textbf{Convergent Anthems}) team has used it as a testbed
\fi
for the paradigm through four linked efforts (Fig.~\ref{fig:casestudy}).

\emph{Computational framework.} The watershed component of the convergence paradigm uses a shared SysML--HFGT ontology to translate physical processes and geospatial data into formal, computationally analyzable models. \citet{Harris:2026:00} first developed simplified SysML watershed reference architectures and instantiated them in HFGT for single- to multi-lake systems, representing fertilizer application and the transport of water and nitrogen from land to lakes and rivers. This approach was extended to the full Chesapeake Bay watershed, translating CAST's geospatial features, including land segments, outlet points, river networks, and estuarine boundaries, into an HFGT representation~\citep{Harris:2026:01}. A weighted least-squares hetero-functional graph state estimator, adapted from work on the American multi-modal energy system~\citep{Thompson:2025:ISC-JR11}, infers unobserved nitrogen and phosphorus flows from partial data while maintaining physical consistency. The resulting framework reproduces CAST's routing assumptions in a transparent, extensible form intended to augment rather than replace it. The same framework has also reproduced economic input--output dynamics, establishing a foundation for coupling regional economic activity with nutrient flows~\citep{naderi:2026:00}.


\emph{Governance and decision support.} The convergence paradigm extends beyond the physical watershed to the institutions responsible for its management. Using the same SysML-based ontology as the watershed model, \citet{Jangjoo:2026:ISC-JR09} represented the Chesapeake Bay Program's (CBP) governance workflow through parallel models of formal rules (rules-in-form) and observed practices (rules-in-use). Overlaying these models reveals where implementation bypasses, supplements, or repurposes prescribed procedures, exposing institutional change mechanisms and the socio-technical dependencies through which they operate. The analysis shows that informal adaptations, including spreadsheet exchanges and preview model runs outside the formal CAST workflow, can help actors navigate resource constraints and political risk while reinforcing disparities in institutional capacity. By making these divergences traceable to qualitative and documentary evidence, SysML provides a formal basis for analyzing governance alongside the physical and computational systems it shapes. Complementary analysis of 29,177 CBP committee documents identifies persistent shifts in governance attention following updates to modeling tools, suggesting that these tools shape not only environmental analysis but also the policy agenda~\citep{jangjoo:2025:01}. Finally, MBSE and HFGT have been applied to resource-constrained project scheduling~\citep{Hosseini:2025:ISC-J55,Hosseini:2025:ISC-JR17}, enabling future extension of the convergence approach to the multi-agency enterprise responsible for implementation in the Chesapeake Bay Watershed region.

\emph{Team science.} Convergence also depends on how the team itself works. Building on the Toolbox Dialogue Initiative and the framework developed by \citet{bennett:2018:00}, \ifanon the project team \else the Convergent Anthems team \fi co-developed and annually revises a Collaboration Agreement setting expectations for communication, authorship and governance, and tracks conflict, trust and cohesion through periodic surveys \citep{fiore:2015:00}. SysML diagrams have become everyday boundary objects within the team, exposing differences in how hydrologists, economists and policy scholars define the same system elements.

\emph{Pedagogy.} The paradigm is taught as well as practiced. 
\ifanon Graduate courses at two universities \else Courses at Virginia Tech and Stevens Institute of Technology \fi introduce SysML and HFGT through project-based modules, from single lakes and input--output models to multi-lake systems and transit, hydrogen and urban carbon challenges, with the meta-cognition map as the organizing frame~\citep{Little:2025:ISC-C84}. A shared laboratory manual for data management, software workflows and documentation serves as a further boundary object. The aim is to train \emph{Anthropocene System Integrators}: researchers and practitioners fluent across the five domains and skilled in translating across disciplinary vocabularies, engaging stakeholders and communicating results \citep{schipfer:2026:00}.

Fig.~\ref{fig:casestudy} also highlights what remains to be done. Although the watershed and governance architectures share an ontology, they are not yet coupled in a single computational model. HFGT capability analysis of the governance network, including measures of centrality, redundancy, and controllability at divergence nodes, also remains future work, as do estuarine and urban extensions. The demonstration therefore establishes that a shared ontology can represent hydrological, economic, institutional, and organizational knowledge, but not yet that their integration improves restoration outcomes.

\section{Implications for Global Environmental Change Research}\label{sec:implications}

Four implications towards global environmental change research follow. First, integration efforts should be redirected upstream. Investment in ever larger coupled models will yield diminishing returns unless the conceptual and visual translations that precede them are formalized in a shared ontology. Funders and research programs could treat reference architectures and ontologies as research outputs in their own right, as valuable as datasets and models.

Second, governance must reside within the system description, not outside it. The Chesapeake Bay work shows that institutional analysis can be expressed in the same language as nutrient transport, and that the models themselves shape governance attention. Representing rules-in-form and rules-in-use alongside biophysical processes offers a route to studying how institutions and environments co-evolve, a core concern of global environmental change research.

Third, convergence requires people as well as methods. Anthropocene System Integrators need training across all five domains and in the cross-cutting skills of translation and engagement. Collaboration agreements, team-science evaluation and shared curricula are part of the infrastructure of convergence, not overheads on it. Convergence programs could budget for them explicitly and evaluate them alongside scientific outputs.

Fourth, methods should be designed for transferability from the outset. Because context dependence is a central barrier, a shared ontology should distinguish general system structures, such as land segments, rivers, actors, and rules, from context-specific parameters and practices. A Chesapeake Bay reference architecture could then be re-instantiated for another basin, making differences in system structure and governance explicit and comparable. Applying the paradigm across regions with varying levels of institutional coordination would test its transferability more rigorously than further work in a single watershed.


The paradigm has limits that further work must address. HFGT and SysML have a learning curve. To lower it, the open-source HFGT Toolbox~\citep{Farid:2024:W09} and its practical handbook~\citep{Farid:2025:ISC-BKR01} are being developed to make the methods easier to use, including streamlining the workflow to run simulations within the toolbox. Translating qualitative governance evidence into formal models also risks flattening context and contestation. Finally, the transfer of the framework beyond the Chesapeake Bay, although underway in energy systems analysis in Australia~\citep{qamar:2026:00} and in mega-project planning~\citep{Hosseini:2025:ISC-J55}, remains to be demonstrated for other socio-environmental systems.